\documentclass[twocolumn,prb,showpacs,floatfix]{revtex4}
\usepackage{amssymb}

\usepackage{amsmath}
\usepackage{epsfig}
\usepackage{graphics}

\input{tcilatex}

\begin{document}

\title{Probing entanglement via Rashba-induced shot noise oscillations\footnote{To appear in the Proceedings of the
NATO ARW   and DARPA Meeting, \textsl{Frontiers of Spintronics and
Optics in Semiconductors: A Symposium in Honor of E. I. Rashba},
Boston, June 2002 (to be published in a special issue of the
Journal of Superconductivity).}}
\author{J. Carlos \surname{Egues}}
\altaffiliation[Permanent address: ]{Department of Physics and Informatics, University of S\~ao Paulo at S\~ao
Carlos, 13560-970 S\~ao Carlos/SP, Brazil.}
\author{Guido Burkard}
\author{Daniel Loss}
\affiliation{Department of Physics and Astronomy, University of Basel, Klingelbergstrasse
82, CH-4056 Basel, Switzerland}

\begin{abstract}
We have recently calculated shot noise for entangled and spin-polarized
electrons in novel beam-splitter geometries with a local Rashba s-o
interaction in the incoming leads. This interaction allows for a
gate-controlled rotation of the incoming electron spins. Here we present an
alternate simpler route to the shot noise calculation in the above work and
focus on only electron pairs. Shot noise for these shows continuous bunching
and antibunching behaviors. In addition, entangled and unentangled triplets
yield distinctive shot noise oscillations. Besides allowing for a direct way
to identify triplet and singlet states, these oscillations can be used to
extract s-o coupling constants through noise measurements. Incoming leads
with spin-orbit interband mixing give rise an additional modulation of the
current noise. This extra rotation allows the design of a spin transistor
with enhanced spin control.
\end{abstract}

\date{\today }
\maketitle

\section{Introduction}

Coherent control of entangled\cite{entangler} and spin-polarized \cite%
{spin-pol,egues} electrons is relevant for the emerging fields of
spintronics \cite{als} and quantum computation.\cite{ghd,als} Spin
rotation of electron states is perhaps the simplest example of
spin manipulation; and yet, it constitutes a relevant operation
for quantum processing and the design of novel spintronic devices.
Spin precession of electrically-injected carriers has recently
been accomplished for diffusive transport in metallic
wires;\cite{jedema} however, this basic operation is still
challenging for transport in ballistic channels.

The Rashba spin-orbit (s-o) interaction,\cite{rashba} present in
quantum-confined heterostructures lacking \emph{structural}
inversion symmetry, offers an interesting possibility to
coherently rotate spin states in the \emph{absence} of a magnetic
field. This was first recognized by Datta and Das \cite{datta} in
their spin-transistor proposal. Essential to this proposal is the
electric control of the s-o interaction in ballistic
one-dimensional channels.\cite{nitta}

We have recently proposed\cite{egd} the use of a\ ``local'' s-o
Rashba interaction as a means of modulating current and shot noise
for spin-polarized and entangled electrons in novel beam-splitter
geometries,\cite{liu,oliver} Fig. 1. The local Rashba interaction
acting within an extension $L$ of lead 1 allows for
gate-controlled spin rotation of the incoming
electrons.\cite{feve} This s-o induced spin rotation produces
continuous changes in the symmetry of the spin part of the wave
function which in turn affects the orbital motion and hence
transport properties such as current and its fluctuations.

\begin{figure}[th]
\begin{center}
\epsfig{file=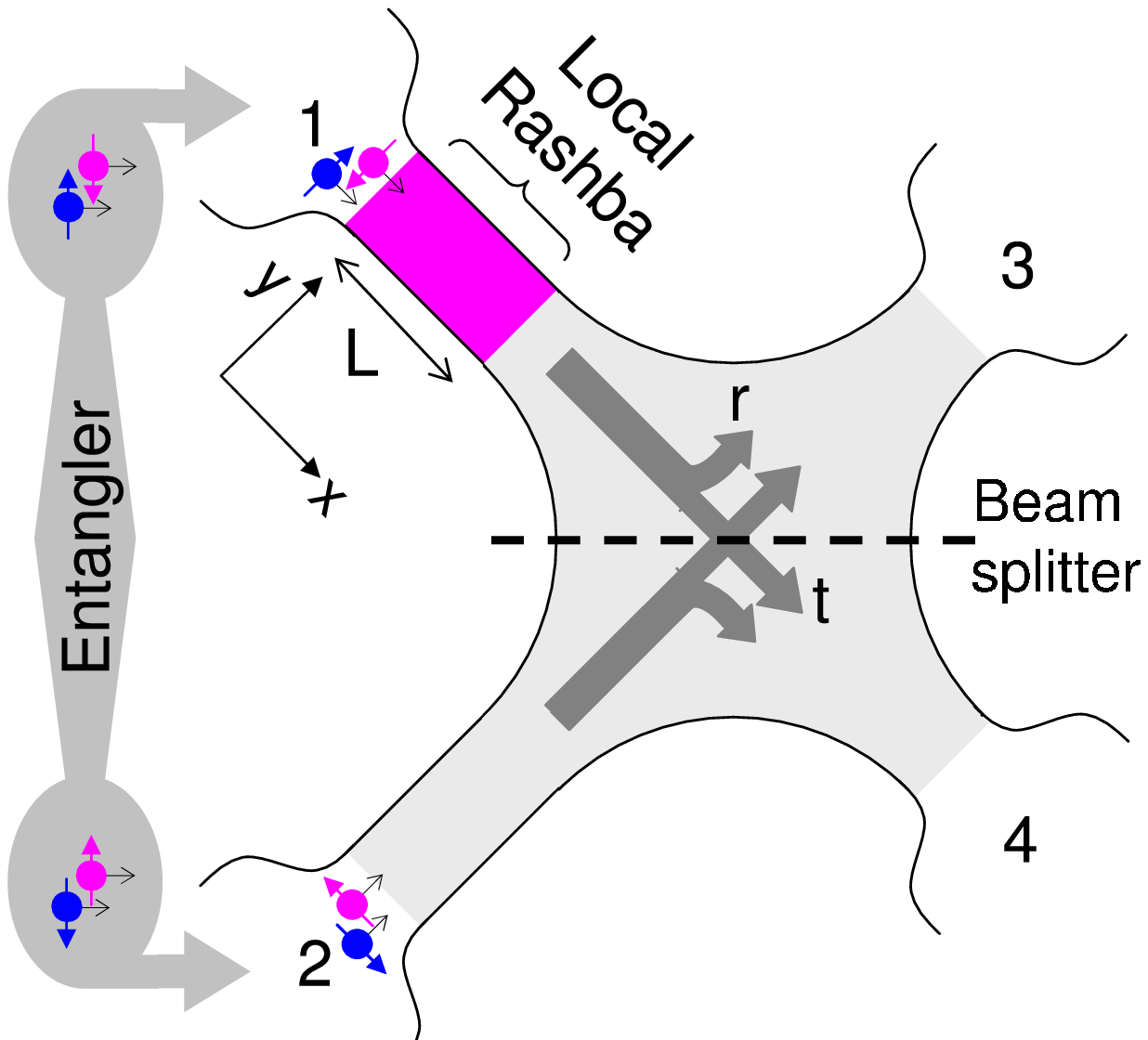, width=0.35\textwidth}
\end{center}
\caption{Beam-splitter geometry with a local Rashba s-o
interaction in lead 1. Entangled pairs are injected in leads 1 and
2. The portion of the entangled pairs traversing lead 1 undergoes
a Rashba-induced spin rotation. This \emph{continuously} changes
the symmetry of the \emph{spin} part of the pair wave function
thus inducing sizable oscillations: continuous bunching and
antibunching behaviors. (Adapted from Ref. [\onlinecite{egd}].)}
\label{fig:fig1}
\end{figure}

Here we present an alternate route to our shot noise calculation
in a beam splitter with a local Rashba interaction.\cite{egd} We
focus on triplet and singlet states only. We determine noise by
first evolving the incoming electron states within lead 1 under
the s-o interaction and then calculating the appropriate noise
matrix elements involving these states. This less general approach
allows us to determine shot noise in the presence of spin orbit
straightforwardly in terms of the earlier results for bunching and
antibunching in a beam-splitter geometry with no spin
orbit.\cite{BLS,taddei} We consider incoming leads with one and
two modes; for the latter we also include a weak s-o induced
interband coupling.\cite{s-o} For entangled singlet and triplet
states, we find \emph{continuous} bunching and antibunching
behavior as a function of the s-o--induced phase. In addition,
our calculation shows that unentangled and entangled triplets display \emph{%
distinctive} shot noise; this enhances the possibilities for
entanglement detection schemes solely based on noise measurements.
Spin-orbit interband mixing induces a further modulation on the
noise for triplets and singlet. This is due to the coherent
transfer of carriers between the s-o coupled Rashba bands as they
traverse lead 1. This additional spin rotation allows the design
of a spin transistor with enhanced spin control.\cite{apl-egues}

This paper is organized as follows. In Sec. II we present our model
Hamiltonian for one-dimensional channels with spin orbit. In Sec. III we
define shot noise, outline our approach, and present results for
beam-splitter configurations with and without s-o induced interband mixing
in the incoming leads. In Sec. VI we briefly describe an extended Datta-Das
transistor with additional spin control due to s-o induced interband mixing.
Section V summarizes our conclusions.

\section{Spin-orbit coupling in 1D channels}

The beam-splitter leads in Fig. 1 are essentially ballistic quantum wires
(``quantum point contacts''). Such one-dimensional channels can be defined
by a gate-induced confining potential $V(y)$ on top of a two dimensional
electron gas
\begin{equation}
H=-\frac{\hbar ^{2}}{2m^{\ast }}\left( \frac{\partial ^{2}}{\partial x^{2}}+%
\frac{\partial ^{2}}{\partial y^{2}}\right) +V(y),  \label{eq1}
\end{equation}%
$m^{\ast }$ is the electron effective mass. The solution to the
corresponding Schr\"{o}dinger equation is
\begin{equation}
\varphi _{k_{x},n,\sigma _{z}}(x,y)=\frac{e^{ik_{x}x}}{\sqrt{L_{x}}}\phi
_{n}(y)|\sigma _{z}\rangle ,  \label{eq2}
\end{equation}%
where $|\sigma _{z}\rangle $ denotes the electron spin component ($\sigma
_{z}$ basis), with eigenenergies%
\begin{equation}
\varepsilon _{k_{x},n,\sigma _{z}}=\frac{\hbar ^{2}k_{x}^{2}}{2m^{\ast }}%
+\epsilon _{n},\text{ }n=a,b...\text{ .}  \label{q3}
\end{equation}%
The transverse confining functions $\phi _{n}(y)$ satisfy%
\begin{equation}
-\frac{\hbar ^{2}}{2m^{\ast }}\frac{d^{2}\phi _{n}(y)}{dy^{2}}+V(y)\phi
_{n}(y)=\epsilon _{n}\phi _{n}(y).  \label{eq4}
\end{equation}%
The confining potential can be chosen as either a parabolic or an
infinite-barrier potential.

\emph{Rashba wire}. Here we consider the Hamiltonian (\ref{eq1}) with the
additional Rashba term\cite{rashba},%
\begin{equation}
H_{R}=i\alpha \left( \sigma _{y}\partial _{x}-\sigma _{x}\partial
_{y}\right) .  \label{eq5}
\end{equation}%
where $\alpha $ is the s-o coupling constant and $\partial _{i}\equiv
\partial /\partial i$, $i=x,y$. In a perturbative fashion, we\ can derive a
reduced Hamiltonian for the problem by expanding the solution of the
corresponding Schr\"{o}dinger equation in the basis of the wire with no s-o $%
\left\{ \varphi _{k_{x},n,\sigma _{z}}(x,y)\right\} $. For two subbands we
find\cite{mireles}
\begin{equation}
H=\left[
\begin{array}{cccc}
\frac{\hbar ^{2}k_{x}^{2}}{2m^{\ast }}+\epsilon _{a} & i\alpha k_{x} & 0 &
-i\alpha d_{ab} \\
-i\alpha k_{x} & \frac{\hbar ^{2}k_{x}^{2}}{2m^{\ast }}+\epsilon _{a} &
-i\alpha d_{ab} & 0 \\
0 & i\alpha d_{ab} & \frac{\hbar ^{2}k_{x}^{2}}{2m^{\ast }}+\epsilon _{b} &
i\alpha k_{x} \\
i\alpha d_{ab} & 0 & -i\alpha k_{x} & \frac{\hbar ^{2}k_{x}^{2}}{2m^{\ast }}%
+\epsilon _{b}%
\end{array}%
\right] ,  \label{eq6}
\end{equation}%
where $d_{ab}\equiv \langle \phi _{a}|\partial /\partial y|\phi _{b}\rangle $
defines the \ `interband coupling' in our system. For infinite-barrier
confinement $d_{ab}=8/3w$, $w$: is the width of the transverse channel. The
Hamiltonian (\ref{eq6}) can be easily diagonalized. Below we discuss the
cases with and without interband coupling.

\begin{figure}[th]
\begin{center}
\epsfig{file=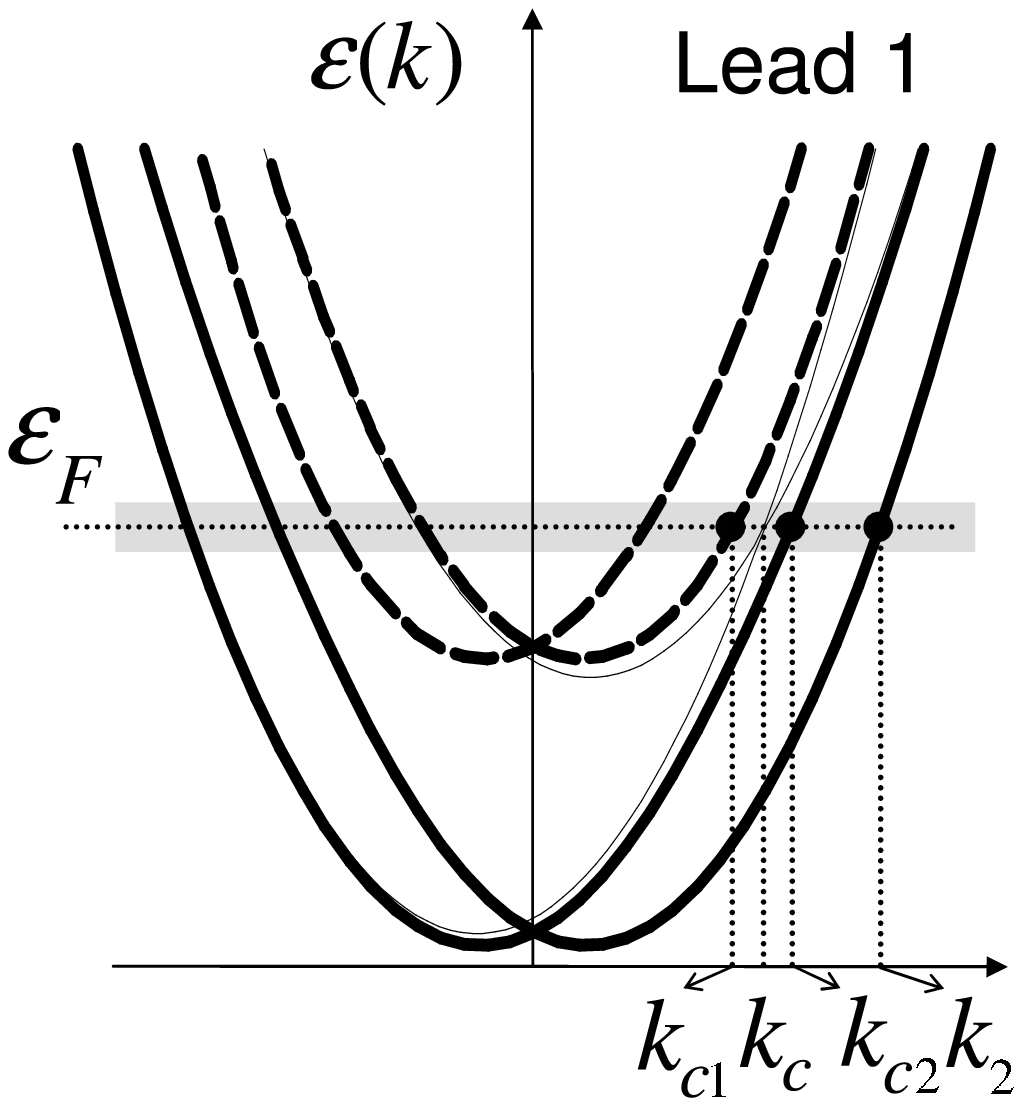, width=0.35\textwidth}
\end{center}
\caption{Two Rashba bands with interband s-o mixing (solid and dashed thick
lines). In absence of interband mixing (solid thin lines) the bands cross at
particular $k$ values (e.g., $k_{c}$). Spin-orbit induced interband coupling
splits up the crossings. Impinging \textit{spin up} electrons with energies $%
\protect\varepsilon _{F}$ near the band crossings (shaded region) are
injected into Rashba states corresponding to the wave vectors $k_{c1}$, $%
k_{c2}$, and $k_{2}$. }
\label{fig:fig2}
\end{figure}

\subsection{Zero interband coupling ($d_{ab}=0$): single spin rotation $%
\protect\theta _{R}$}

Here the problem is strictly one dimensional; this should be a
good approximation for $\alpha d_{ab}$ much smaller than the
interband separation. In this case the solution is well
known:\cite{mireles} two uncoupled Rashba subbands since the
Hamiltonian (\ref{eq6}) is block diagonal for $d_{ab}=0$
\begin{equation}
\varepsilon _{a,b}^{(s)}(k_{x})=\frac{\hbar ^{2}}{2m^{\ast }}\left(
k_{x}-sk_{R}\right) ^{2}+\epsilon _{a,b}-\frac{\hbar ^{2}k_{R}^{2}}{2m^{\ast
}}\text{, }s=\pm ,  \label{eq7}
\end{equation}%
where $k_{R}\equiv $ $m^{\ast }\alpha /\hbar ^{2}$ is the Rashba wave
vector. Note that in the absence of s-o induced interband coupling, the
Rashba subbands cross,\ Fig. 2.\ The crossing for positive wave vectors
occurs at
\begin{equation}
k_{c}=\frac{\epsilon _{b}-\epsilon _{a}}{2\alpha },  \label{eq7a}
\end{equation}%
which we obtain from $\varepsilon _{a}^{-}(k_{c})=\varepsilon
_{b}^{+}(k_{c}).$

Spin-up electrons injected into either subband of the wire spin precess as
they move down the channel. As first discussed by Datta and Das, a spin up
electron traversing the length $L$ of the Rashba region evolves to%
\begin{equation}
|\uparrow \rangle \rightarrow \cos \theta _{R}/2|\uparrow \rangle +\sin
\theta _{R}/2|\downarrow \rangle \text{,}  \label{eq8}
\end{equation}%
where $\theta _{R}=2m^{\ast }\alpha L/\hbar ^{2}$ is the precessing angle
about the $-y$ axis. The above assumes a unity transmission\cite{mismatch}
across the Rashba region.

We are particularly interested in incoming two-particle states, e.g.,
triplets and singlet pairs injected in leads 1 and 2, Fig. 1. The portion of
these electron pairs crossing lead 1 undergoes a Rashba spin rotation. For
instance, incoming singlet ($-$) and triplet ($+$) states yield
\begin{eqnarray}
|\uparrow \downarrow \rangle \mp |\downarrow \uparrow \rangle \rightarrow
&&\cos \theta _{R}/2[|\uparrow \downarrow \rangle \mp |\downarrow \uparrow
\rangle ]+  \notag \\
&&\sin \theta _{R}/2[|\downarrow \downarrow \rangle \pm |\uparrow \uparrow
\rangle ].  \label{eq9}
\end{eqnarray}%
where we use the notation $|\uparrow \uparrow \rangle \equiv |\uparrow
_{1}\uparrow _{2}\rangle $ to denote the electron spin state in the
corresponding leads. Later on we will use the above states to determine the
noise for injected spin entangled pairs. Equation (\ref{eq9}) clearly shows
a change in the symmetry of the spin part of the wave function of the pair:
initially entangled triplet and singlet states at the entrance of the Rashba
region emerge of it with s-o induced mixed in components of unentangled
triplet states.

\subsection{Non-zero interband coupling ($d_{ab}\neq 0$): additional spin
rotation $\protect\theta _{d}$}

In the presence of interband coupling the subband crossings in
Fig. 2 split up, i.e., near the crossing point $k_{c}$ the Rashba
subbands anti cross. This is more easily seen by rewriting the
Hamiltonian (\ref{eq6}) in the basis of the $d_{ab}=0$ problem.
Instead of diagonalizing the problem exactly, here we follow a
more intuitive perturbative approach similar to that of the
nearly-free electron model.\cite{am,apl-egues} For small interband
couplings, we can approximate (lowest order) the energy
dispersions near the crossing at $k_{c}$ (Fig. 2) by%
\begin{equation}
\varepsilon _{\pm }(k)=\frac{\hbar ^{2}k^{2}}{2m}+\frac{1}{2}\epsilon _{a}+%
\frac{1}{2}\epsilon _{b}\pm \alpha d,  \label{eq10}
\end{equation}%
with corresponding zeroth-order spinors ,
\begin{equation}
|\psi _{\pm }\rangle =\frac{1}{\sqrt{2}}[|-\rangle _{a}\pm |+\rangle _{b}]=%
\frac{1}{2}\left(
\begin{array}{c}
1 \\
-i%
\end{array}%
\right) _{a}\pm \frac{1}{2}\left(
\begin{array}{c}
1 \\
i%
\end{array}%
\right) _{b},  \label{eq11}
\end{equation}%
where the subindices \emph{a} and \emph{b} denote the respective Rashba
subbands.

How does an injected spin-up state with energy near the crossing gets
rotated as it crosses the Rashba region? In analogy to the usual Datta-Das
description for the case with no s-o interband coupling, we expand the
incoming spin up state as
\begin{eqnarray}
| &\uparrow &\rangle _{a}e^{ikx}|_{x\rightarrow 0^{-}}=  \notag \\
&&\frac{1}{\sqrt{2}}\left\{ \frac{1}{\sqrt{2}}\left[ |\psi _{+}\rangle
e^{ik_{c1}x}+|\psi _{-}\rangle e^{ik_{c2}x}\right] +|+\rangle
_{a}e^{ik_{2}x}\right\} _{x\rightarrow 0^{+}},  \label{eq12}
\end{eqnarray}%
where $x=0$ defines the entrance of the Rashba
region.\cite{mismatch} Note that in the above expansion we only
include \emph{three} of the four
intersection points in Fig. 2, $k_{c1}=k_{c}-\Delta /2$, $%
k_{c2}=k_{c}+\Delta /2$, with $\Delta =\theta _{R}d_{ab}/k_{c}$, and $k_{2}$
. The parameter $\Delta $ is obtained by imposing $\varepsilon
_{+}(k_{c1})=\varepsilon _{-}(k_{c2})$ near the crossing point $k_{c}$,
i.e., for energies approximately equal to $\varepsilon _{a}^{-}(k_{c})=\hbar
^{2}k_{c}^{2}/2m^{\ast }+\alpha k_{c}+\epsilon _{a}$. By substituting $|\psi
_{\pm }\rangle $ [Eq. (\ref{eq11})] into Eq. (\ref{eq12}) we can easily see
that the boundary condition for the wave function is satisfied at $x=0.$ It
is straightforward to verify that the boundary condition for the proper
derivative of the wave function\cite{erasmo} is also satisfied, provided
that $\Delta <<4k_{F}$; this condition is well satisfied for the parameters
we use in our calculation. By following the same reasoning as for the
uncoupled case Sec. (II-A), we find that an impinging spin-up electron in
channel \emph{a }evolves to the state
\begin{eqnarray}
|\uparrow \rangle _{a} &\rightarrow &\frac{1}{2}\left[ \cos (\theta
_{d}/2)e^{-i\theta _{R}/2}+e^{i\theta _{R}/2}\right] |\uparrow \rangle _{a}+
\notag \\
&&\frac{1}{2}\left[ -i\cos (\theta _{d}/2)e^{-i\theta _{R}/2}+ie^{i\theta
_{R}/2}\right] |\downarrow \rangle _{a}-  \notag \\
&&\frac{1}{2}i\sin (\theta _{d}/2)e^{-i\theta _{R}/2}|\uparrow \rangle _{b}+
\notag \\
&&\frac{1}{2}\sin (\theta _{d}/2)e^{-i\theta _{R}/2}|\downarrow \rangle _{b},
\label{eq13}
\end{eqnarray}%
upon crossing the Rashba region of length $L$ in lead 1. Equation (\ref{eq13}%
) clearly shows an additional modulation $\theta _{d}$ for electrons
impinging near the band crossing,\ Fig. 2. Note that the above state vector
yields the usual Datta-Das state (\ref{eq8}) for $\theta _{d}=0$. The extra
spin-rotation $\theta _{d}=\theta _{R}d/k_{c}$ can, in principle, be varied
independently of $\theta _{R}$ via proper lateral gating structures.

In the presence of s-o interband mixing, the portion of an electron pair
crossing lead 1 undergoes a rotation $\theta _{d}$ in addition to the usual
rotation $\theta _{R}$. For instance, upon crossing the Rashba region in
lead 1 an unentangled spin-up triplet emerges as
\begin{eqnarray}
|\uparrow \uparrow \rangle _{a} &\rightarrow &\frac{1}{2}\left[ \cos (\theta
_{d}/2)e^{-i\theta _{R}/2}+e^{i\theta _{R}/2}\right] |\uparrow \uparrow
\rangle _{aa}+  \notag \\
&&\frac{1}{2}\left[ -i\cos (\theta _{d}/2)e^{-i\theta _{R}/2}+ie^{i\theta
_{R}/2}\right] |\downarrow \uparrow \rangle _{aa}-  \notag \\
&&\frac{1}{2}i\sin (\theta _{d}/2)e^{-i\theta _{R}/2}|\uparrow \uparrow
\rangle _{ba}+  \notag \\
&&\frac{1}{2}\sin (\theta _{d}/2)e^{-i\theta _{R}/2}|\downarrow \uparrow
\rangle _{ba},  \label{eq14}
\end{eqnarray}%
where we use the notation $|\downarrow \uparrow \rangle _{ba}\equiv $ $%
|\downarrow _{1b}\uparrow _{2a}\rangle $. Similar ``evolved'' pairwise
states can be obtained for the singlet and the other triplets.

\subsection{Realistic parameters}

Before we proceed to determine shot noise, let us estimate the size of the
spin rotations $\theta _{R}$ and $\theta _{d}$. For the sake of simplicity
and concreteness, let us assume an infinite (transverse) confining potential
of width $w$. In this case, the s-o interband mixing matrix element is $%
d=8/3w$ and $\epsilon _{b}=3\pi ^{2}\hbar ^{2}/2mw^{2}$\ ($\epsilon
_{a}\equiv 0$, arbitrary origin). The Rashba energy $\epsilon _{R}\equiv
\hbar ^{2}k_{R}^{2}/2m^{\ast }$ defines a natural scale in our system.
Choosing $\epsilon _{b}=16\epsilon _{R}$, we find $\alpha =(\sqrt{3}\pi
/4)\hbar ^{2}/mw=3.45\times 10^{-11}$ eVm (and $\epsilon _{R}\sim 0.39$ meV)
for $m=0.05m_{0}$ (see Ref. [\onlinecite{nitta}]) and $w=60$ nm. In
addition, the energy at the band crossing is $\varepsilon
_{a}^{-}(k_{c})=24\varepsilon _{R}\sim 9.\,\allowbreak 36$ meV; the Fermi
energy should be tuned around this value. The wave vector at the crossing
follows from Eq. (\ref{eq7a}), $k_{c}=8\epsilon _{R}/\alpha $. A
conservative estimate for the spin-rotation angles is obtained by assuming $%
L=69$ nm; this yields $\theta _{R}=\pi $ and $\theta _{d}=\theta
_{R}d/k_{c}=\pi /2$ since $d/k_{c}\sim 0.5$. In addition, we find $\Delta
/4k_{F}\sim 0.05$ for the above parameters which assures that the bounday
condition for the velocity operator is satisfied in our system [see Sec.
(II-B)]. We emphasize the spin-rotation angles $\theta _{R}$ and $\theta
_{d} $ can be independently varied. For instance, proper (side) gates can
induce changes in the width $w$ of the confining potential. Our simple
estimates show that sizable spin rotations should be attainable
experimentally.

\section{Scattering approach for current and noise: basics}

\emph{Definition.} The discreteness of non-equilibrium charge flow gives
rise to intrinsic fluctuations in the electric current: shot noise.
Mathematically, shot noise is defined by the power spectral density of the
current-current autocorrelation function.\ In a multi-lead geometry, the
noise between leads $\gamma $ and $\mu $ is
\begin{equation}
S_{\gamma \mu }(\omega )=\frac{1}{2}\int \langle \delta \hat{I}_{\gamma
}(t)\delta \hat{I}_{\mu }(t^{\prime })+\delta \hat{I}_{\mu }(t^{\prime
})\delta \hat{I}_{\gamma }(t)\rangle e^{i\omega t}dt,  \label{eq15}
\end{equation}%
where $\delta \hat{I}_{\gamma }(t)$ denotes the current fluctuation about
its average in lead $\gamma $ at time $t$. In the Landauer-B\"{u}ttiker
approach\cite{buttiker} the current in lead $\gamma $ reads
\begin{eqnarray}
\hat{I}_{\gamma }(t) &=&\frac{e}{h}\sum_{\alpha \beta }\!\!\int
\!\!d\varepsilon d\varepsilon ^{\prime }e^{i(\varepsilon -\varepsilon
^{\prime })t/\hbar }\mathbf{a}_{\alpha }^{\dagger }(\varepsilon )\mathbf{A}%
_{\alpha ,\beta }(\gamma ;\varepsilon ,\varepsilon ^{\prime })\mathbf{a}%
_{\beta }(\varepsilon ^{\prime }),  \notag \\
&&\mathbf{A}_{\alpha \beta }(\gamma ;\varepsilon ,\varepsilon ^{\prime
})=\delta _{\gamma \alpha }\delta _{\gamma \beta }\mathbf{1}-\mathbf{s}%
_{\gamma \alpha }^{\dagger }(\varepsilon )\mathbf{s}_{\gamma \beta
}(\varepsilon ^{\prime }),  \label{eq16}
\end{eqnarray}%
where $\mathbf{s}$ denotes the scattering matrix of the system and $\mathbf{a%
}_{\alpha }^{\dagger }(\varepsilon )$ and $\mathbf{a}_{\alpha }(\varepsilon
) $ are the usual fermionic creation and annihilation (two-component)
operators for electrons; later on we write these more explicitly in terms of
their spin components.

\emph{Beam splitter }\textbf{s} \emph{matrix.} The relevant scattering
matrix in our problem is that of the beam splitter. We assume the beam
splitter transmits electrons from leads 1 to 4 and from leads 2 to 3 with an
amplitude $t$ and from leads 1 to 3 and from 2 to 4 with an amplitude $r$.
Hence
\begin{equation}
\mathbf{s}=\left(
\begin{array}{cccc}
0 & 0 & \mathbf{s}_{13} & \mathbf{s}_{14} \\
0 & 0 & \mathbf{s}_{23} & \mathbf{s}_{24} \\
\mathbf{s}_{31} & \mathbf{s}_{32} & 0 & 0 \\
\mathbf{s}_{41} & \mathbf{s}_{42} & 0 & 0%
\end{array}%
\right) =\left(
\begin{array}{cccc}
0 & 0 & r & t \\
0 & 0 & t & r \\
r & t & 0 & 0 \\
t & r & 0 & 0%
\end{array}%
\right) .  \label{q17}
\end{equation}%
Note that backscattering into the incoming leads is neglected in $\mathbf{s}$%
. In addition, $\mathbf{s}$ is assumed to be both spin \emph{and} channel
independent.

In what follows we determine shot noise for electron pairs. In this case the
angle brackets in the noise definition [Eq. (\ref{eq15})] should be
interpreted as a\ quantum mechanical expectation value between entangled or
unentangled electron states.

\section{Shot noise in the absence of spin orbit}

\subsection{Bunching and antibunching for entangled electrons: earlier
results}

Shot noise for singlet and triplets in a beam-splitter geometry with no
spin-orbit interaction was first investigated in Ref. [\onlinecite{BLS}]
within the scattering approach. These authors calculated the expectation
value of the noise between\ singlet\cite{delay}

\begin{equation}
|S\rangle =\frac{1}{\sqrt{2}}\left[ a_{1\uparrow }^{\dagger }(\varepsilon
_{1})a_{2\downarrow }^{\dagger }(\varepsilon _{2})-a_{1\downarrow }^{\dagger
}(\varepsilon _{1})a_{2\uparrow }^{\dagger }(\varepsilon _{2})\right]
|0\rangle \text{,}  \label{eq18}
\end{equation}%
entangled triplet
\begin{equation}
|Te\rangle =\frac{1}{\sqrt{2}}\left[ a_{1\uparrow }^{\dagger }(\varepsilon
_{1})a_{2\downarrow }^{\dagger }(\varepsilon _{2})+a_{1\downarrow }^{\dagger
}(\varepsilon _{1})a_{2\uparrow }^{\dagger }(\varepsilon _{2})\right]
|0\rangle ,  \label{eq19}
\end{equation}%
and unentangled triplets%
\begin{equation}
|Tu_{\sigma }\rangle =a_{1\sigma }^{\dagger }(\varepsilon _{1})a_{2\sigma
}^{\dagger }(\varepsilon _{2})|0\rangle ,\text{ }\sigma =\uparrow
,\downarrow \text{,}  \label{eq20}
\end{equation}%
where $|0\rangle $ denotes the filled Fermi sea of the contacts and $%
a_{\alpha \sigma }^{\dagger }(\varepsilon _{\alpha })$ $[a_{\alpha \sigma
}(\varepsilon _{\alpha })]$ the creation (annihilation) operator for an
incoming electron with energy $\varepsilon _{\alpha }$ in lead $\alpha $; $%
\sigma $ is the spin component along the quantization axis $z$. At zero
temperature, zero frequency, and no applied voltage, the Fermi sea is
completely noiseless and shot noise in the beam-splitter geometry is solely
due to the injected pairs. The noise in lead 3 is found to be [%
\onlinecite{BLS}]
\begin{equation}
S_{33}^{S/Te,u_{\sigma }}=\frac{2e^{2}}{h\nu }T(1-T)(1\pm \delta
_{\varepsilon _{1},\varepsilon _{2}}),  \label{eq21}
\end{equation}%
where $T=|t|^{2}$ denotes the transmission probability through the beam
splitter and $\nu $ the density of states (discrete spectrum) in the leads.
In Eq. (\ref{eq21}) the upper sign refers to singlet and the lower one to
triplet states. Note that all three triplets yield the \emph{same} noise
power.

The important result embodied in (\ref{eq21}) is that singlet and triplet
pairs give rise to shot noise bunching and antibunching, respectively. This
terminology means that shot noise is enhanced for a singlet pair (bunching)
while suppressed for triplets (antibunching) as compared to the shot noise
for uncorrelated particles (``full shot noise''). The Fano factor in lead 3
is given by $F=S_{33}^{S/T}/2eI_{3}=T(1-T)(1\pm \delta _{\varepsilon
_{1},\varepsilon _{2}})$, where $I_{\alpha }=e/h\nu $ defines the average
current in lead $\alpha $. For uncorrelated particles the Fano factor is $%
F_{unc}$ $=T(1-T)$. Note that the Fano factor for a singlet state is
enhanced by a factor of two with respect to that of uncorrelated electrons
while that of triplets is supressed to zero. As first pointed out in\ Ref. [%
\onlinecite{BLS}], this offers the possibility of detecting entanglement via
shot noise measurements. As we discuss below, in the presence of a local
Rashba interaction in lead 1 not only entangled states but also unentangled
triplets can be distinguished via noise measurements.

\section{Shot noise in the presence of spin orbit}

The alternate route to the shot noise calculation we present below is
simpler, though less general, than that of Ref. [\onlinecite{egd}] in which
an extended spin-dependent scattering formalism was used. Here we
essentially evaluate the matrix element in the noise definition (\ref{eq15})
by considering the appropriate Rashba-rotated states for one channel and two
channels with s-o interband mixing.

\subsection{One subband case: single modulation $\protect\theta _{R}$}

For a single channel the relevant entangled states to the noise calculation
are those in Eq. (\ref{eq9}). Below we rewrite these in terms the singlet
and triplet states in Eqs. (\ref{eq18})-(\ref{eq20})%
\begin{equation}
|S/T_{e}\rangle _{L}=\cos (\theta _{R}/2)|S/Te\rangle +\frac{1}{\sqrt{2}}%
\sin (\theta _{R}/2)[|Tu_{\downarrow }\rangle \pm |Tu_{\uparrow }\rangle ],
\label{eq23}
\end{equation}%
the above state clearly shows that the Rashba s-o interaction mixes up
entangled and unentangled pairs as they cross the Rashba region of length $L$%
. Since the beam splitter $\mathbf{s}$ matrix is spin independent, the shot
noise in lead 3 corresponding to $|S/T_{e}\rangle _{L}$ is simply%
\begin{equation}
S_{s-o}^{S/Te_{z}}(\theta _{R})=\cos ^{2}(\theta _{R}/2)S_{33}^{S/Te}+\frac{1%
}{2}\sin ^{2}(\theta _{R}/2)(S_{33}^{Tu_{\downarrow }}\pm
S_{33}^{Tu_{\uparrow }}),  \label{eq24}
\end{equation}%
i.e., the noise in the presence of spin orbit can be expressed in terms of
the earlier results with no spin orbit interaction, Sec. (IV-A).
Substituting Eq. (\ref{eq21}) in the above we find%
\begin{eqnarray}
S_{s-o}^{S/Te_{z}}(\theta _{R}) &=&\frac{2e^{2}}{h\nu }T(1-T)[\cos
^{2}(\theta _{R}/2)(1\pm \delta _{\varepsilon _{1},\varepsilon _{2}})+
\notag \\
&&\sin ^{2}(\theta _{R}/2)(1-\delta _{\varepsilon _{1},\varepsilon _{2}})],
\label{eq25}
\end{eqnarray}%
which for the singlet simplifies to
\begin{equation}
S_{s-o}^{S}(\theta _{R})=\frac{2e^{2}}{h\nu }T(1-T)[1+\cos (\theta
_{R})\delta _{\varepsilon _{1},\varepsilon _{2}}],  \label{eq26}
\end{equation}%
and for the entangled triplet
\begin{equation}
S_{s-o}^{Te_{z}}(\theta _{R})=\frac{2e^{2}}{h\nu }T(1-T)(1-\delta
_{\varepsilon _{1},\varepsilon _{2}}).  \label{eq27}
\end{equation}%
Note that the entangled singlet and triplet states are labelled with respect
to the \emph{z} axis. More symmetrical formulas can be obtained for a
quantization axis along the Rashba rotation axis ($-$\emph{y}) (see Ref. [%
\onlinecite{egd}]).

Similarly, the unentangled triplets along the \emph{z} axis yield
\begin{equation}
S_{s-o}^{Tu_{\uparrow }}(\theta _{R})=S_{s-o}^{Tu_{\downarrow }}(\theta
_{R})=\frac{2e^{2}}{h\nu }T(1-T)[1-\cos ^{2}(\theta _{R}/2)\delta
_{\varepsilon _{1},\varepsilon _{2}}].  \label{eq27a}
\end{equation}%
Interestingly enough, the entangled and unentangled triplets along the \emph{%
z} direction, Eqs. (\ref{eq27}) and (\ref{eq27a}), display distinct shot
noise as a function of the Rashba angle $\theta _{R}$. This is in contrast
to the case with no local Rashba interaction for which all triplets show
identical noise [cf. Eq. (\ref{eq21}), Sec. (IV-A)]. This feature makes it
possible to distinguish unentangled from entangled\ triplet states via noise
measurements as a function of the Rashba phase.


\subsection{Two subbands + interband mixing: additional modulation $\protect%
\theta _{d}$}

Here the portion of the injected pairs propagating in lead 1 undergoes both
the ordinary Rashba rotation $\theta _{R}$ and the additional rotation $%
\theta _{d}$ due to interband coupling, Eq. (\ref{eq13}). To determine shot
noise in this case we proceed following the straightforward calculation in
the preceding section; here, however, we use generalized spin-rotated
states. For the spin-up unentangled triplet in Eq. (\ref{eq14}), we find for
the noise in lead 3%
\begin{eqnarray}
&&S_{s-o}^{Tu_{\uparrow }}(\theta _{R},\theta _{d}) =\frac{2e^{2}}{h\nu }%
T(1-T)  \notag \\
&&\left[ 1-\frac{1}{2}\left( 1-\frac{1}{2}\sin ^{2}\left( \theta
_{d}/2\right) +\cos \left( \theta _{d}/2\right) \cos \theta _{R}\right)
\delta _{\varepsilon _{1},\varepsilon _{2}}\right] .\text{ }  \label{eq28}
\end{eqnarray}%
Note that $S_{s-o}^{Tu_{\uparrow }}(\theta _{R},\theta
_{d})=S_{s-o}^{Tu_{\downarrow }}(\theta _{R},\theta _{d}).$ Similarly, for
entangled triplet and singlet states we find%
\begin{equation}
S^{Te_{z}}(\theta _{R},\theta _{d})=\frac{2e^{2}}{h\nu }T(1-T)\left[ 1-\frac{%
1}{2}\left( \cos ^{2}\left( \theta _{d}/2\right) +1\right) \delta
_{\varepsilon _{1},\varepsilon _{2}}\right] ,  \label{eq29}
\end{equation}%
and

\begin{equation}
S^{S}(\theta _{R},\theta _{d})=\frac{2e^{2}}{h\nu }T(1-T)\left[ 1+\left(
\cos (\theta _{d}/2)\cos \theta _{R}\right) \delta _{\varepsilon
_{1},\varepsilon _{2}}\right] ,  \label{eq30}
\end{equation}%
respectively. Note the additional modulation $\theta _{d}$ due to s-o
interband mixing in the above expressions. For $\theta _{d}=0$ these reduce
to the previous case [Eqs. (\ref{eq26}) and (\ref{eq27})].


\subsection{Some plots of the noise modulation: Fano factors}

The average current in the incoming leads is $I_{1}=I_{2}=2e/h\nu $. By
normalizing the shot noise expressions (\ref{eq28}), (\ref{eq29}), and (\ref%
{eq30}) by $2eIT(1-T)$ we obtain the ``reduced'' Fano factors%
\begin{eqnarray}
f_{Tu_{\uparrow }} &=&1-\frac{1}{2}\left( 1-\frac{1}{2}\sin ^{2}\left(
\theta _{d}/2\right) +\cos \frac{\theta _{d}}{2}\cos \theta _{R}\right)
\delta _{\varepsilon _{1}\varepsilon _{2}},  \label{eq30a} \\
f_{Te_{z}} &=&1-\frac{1}{2}\left( \cos ^{2}\frac{\theta _{d}}{2}+1\right)
\delta _{\varepsilon _{1}\varepsilon _{2}},  \label{eq30b} \\
f_{S} &=&1+\left( \cos \frac{\theta _{d}}{2}\cos \theta _{R}\right) \delta
_{\varepsilon _{1}\varepsilon _{2}}.  \label{eq30c}
\end{eqnarray}%
Here, again, the triplets $f_{Te_{z}}$ and $f_{Tu_{\uparrow
}}=f_{Tu_{\downarrow }}$ show distinctive noise; the additional modulation $%
\theta _{d}$ makes them even more distinct [cf. Eqs. (\ref{eq30b}) and (\ref%
{eq30c}) to (\ref{eq26}) and (\ref{eq27})]. Figure 3 illustrates the angular
dependence of the Fano factor $f_{S}$ for a singlet pair. Note the
continuous bunching and antibunching as a function of the angles $\theta
_{R} $ and $\theta _{d}$; these angles are, in principle, independently
tunable via a proper gating structure. The Fano factor for an entangled
triplet pair along the Rashba rotation axis ($-y$) displays a dependence
similar to that in Fig. 3, i.e., $f_{Te_{y}}=1-\left( \cos \frac{\theta _{d}%
}{2}\cos \theta _{R}\right) \delta _{\varepsilon _{1}\varepsilon _{2}}$.

\begin{figure}[th]
\begin{center}
\epsfig{file=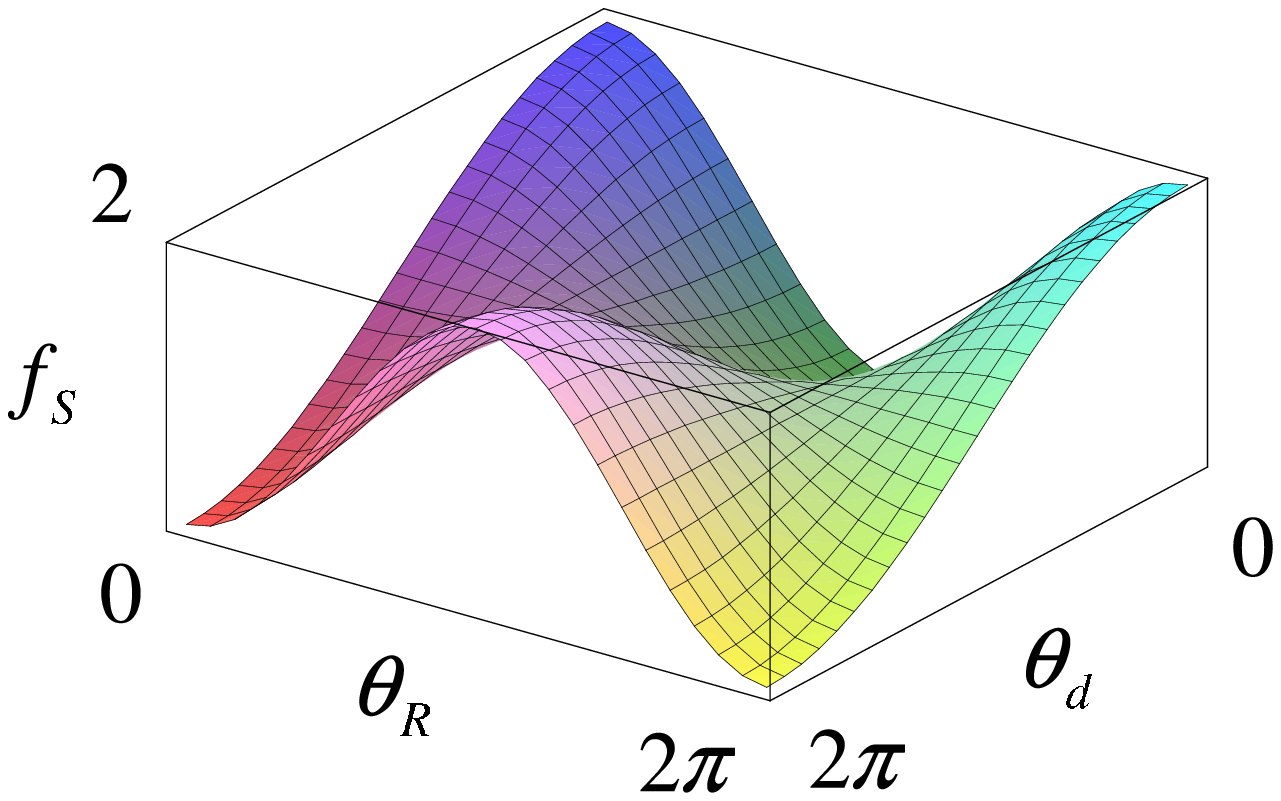, width=0.37\textwidth}
\end{center}
\caption{Angular dependence of the Fano factor $f_S$ for a single pair.}
\label{fig:fig3}
\end{figure}

\subsection{Extracting s-o\ coupling constants via noise measurements}

The sizable Rashba-induced oscillations in the Fano factor plotted in\ Fig.
3 allows for a direct determination of s-o coupling constants via noise
measurements since
\begin{equation}
\alpha =\frac{\hbar ^{2}}{m^{\ast }L}\arccos \sqrt{f_{S}},  \label{eq31}
\end{equation}%
from Eq. (\ref{eq30a})\ for $\theta _{d}=0$.

\section{A new ``spin'' on Datta-Das transistor?}

\emph{Electrons impinging near the crossing.} The additional spin rotation $%
\theta _{d}$ due to the transfer of electrons between s-o coupled bands
allows for the design of a spin transistor with enhanced capabilities, i.e.,
with extra spin control. Let us consider a two-terminal geometry with
spin-polarized emitter (source) and collector (drain) and a
quasi-one-dimensional wire connecting them. We consider a wire with
two-subbands in the presence of s-o induced interband coupling. As explained
in Sec. (II-B), electrons transversing the wire undergo spin rotations $%
\theta _{R}$ and $\theta _{d}$ due to the Rashba interaction itself and\ the
s--o induced interband coupling. Neglecting backscattering in the wire, we
can easily determine the spin-resolved current from Eq. (\ref{eq13}).
Assuming a spin-up source, we find\cite{apl-egues}%
\begin{equation}
I_{\uparrow ,\downarrow }\varpropto 1\pm \cos (\theta _{d}/2)\cos (\theta
_{R}).  \label{eq32}
\end{equation}%
For $\theta _{d}=0$ Eq. (\ref{eq32}) yields the usual current modulation in
the Datta-Das transistor. We emphasize that the additional modulation $%
\theta _{d}$ occurs only for electrons with energies near the band crossing
at $k_{c}$, see shaded region in Fig. 2. Electrons away from the crossing
undergo the single rotation $\theta _{R}$.

\section{Conclusion}

A local Rashba spin-orbit interaction can strongly modulate both current and
shot noise in novel beam-splitter geometries. For entangled electron pairs
this interaction leads to tunable (continuous) bunching and antibunching
behavior. Triplets states display distinctive noise as a function of the
Rashba-induced phase; this allows one to distinguish triplets (entangled
from unentangled) via noise measurements. These measurements can also
extract spin-orbit coupling constants. Spin-orbit induced interband coupling
in the incoming leads gives rise to additional spin rotation. This further
modulates noise and current for electrons impinging near the band crossings.
This feature should be relevant for the design of a spin-transistor with
additional spin control. Our simple estimates indicate that sizable spin
rotations are achievable for realistic system parameters.

This work was supported by NCCR\ Nanoscience, the Swiss NSF, DARPA, and ARO.
We acknowledge useful discussions with C. Schroll, H. Gassmann, and D.
Saraga.


\begin{thebibliography}{99}
\bibitem{entangler} See P. Recher \textit{et al.} [Phys.\ Rev.\ B \textbf{64}%
, 165314 (2001)] and D. Saraga and D. Loss (cond-mat/0205553) for
theoretical proposals of electron entanglers using quantum dots.

\bibitem{spin-pol} R.\ Fiederling \textit{et al.}, Nature \textbf{402}, 787
(1999); Y. Ohno \textit{et al.,} Nature \textbf{402}, 790 (1999).

\bibitem{egues} See J. C. Egues [Phys.\ Rev.\ Lett.\ \textbf{80}, 4578
(1998)] and J. C. Egues \emph{et al. }[Phys. Rev. B \textbf{64}, 195319
(2001)] for \emph{ballistic }spin filtering in semimagnetic heterostructures.

\bibitem{als} {\it Semiconductor Spintronics and Quantum
Computation}, Eds. D. D. Awschalom, D. Loss, and N. Samarth
(Springer, Berlin, 2002).

\bibitem{ghd} G.\ Burkard \emph{et al.}, Fortschr. Physc. \textbf{48}, 965
(2000); see also Ref. [\onlinecite{als}].

\bibitem{jedema} F. J. Jedema \emph{et al.}, Nature \textbf{416}, 713 (2002).

\bibitem{rashba} Yu.\ A. Bychkov and E. I. Rashba, JETP Lett.\ \textbf{39},
78 (1984).

\bibitem{datta} S. Datta and B. Das, Appl.\ Phys. Lett.\ \textbf{56}, 655
(1990). See G. Meir \emph{et al.} [Phys. Rev. B \textbf{65}, 125327 (2002)]
and C.-M. Hu \emph{et al.} [J. Appl.\ Phys. \textbf{91}, 7251 (2002)] for a
description of current experimental efforts for the realization of the
Datta-Das transistor.

\bibitem{nitta} Gate control of the s-o constant has been achieved in 2DEGs;
see G. Engels \emph{et al.} Phys.\ Rev.\ B \textbf{55}, R1958
(1997) and J. Nitta \emph{et al.}, Phys.\ Rev.\ Lett.\
\textbf{78}, 1335 (1997). See also D. Grundler [Phys.\ Rev.\
Lett.\ \textbf{84}, 6074 (2000)] for gate control using additional
back gates to keep the carrier concentration constant. For a
recent experiment measuring the Rashba coupling constant via
``weak antilocalization analysis'', see T. Koga {\emph et al.}
Phys. Rev. Lett. {\bf 89}, 046801 (2002).

\bibitem{egd} J.\ C. Egues, G. Burkard, and D. Loss, cond-mat/0204639 (submitted).

\bibitem{liu} R. C. Liu \textit{et al.}, Nature (London), \textbf{391}, 263
(1998).

\bibitem{oliver} W. D. Oliver \textit{et al.}, in \textit{Quantum Mesoscopic
Phenomena and Mesoscopic Devices in Microelectronics}, vol. 559 of NATO ASI
Series C: Mathematical and Physical Sciences, eds. I. O. Kulik and R.
Ellialtioglu (Kluwer, Dordrecht, 2000), pp. 457-466.

\bibitem{feve} See G. Feve \emph{et al.} (cond-mat/0108021) for a
formulation of the scattering formalism in terms of the Rashba states for a
beam-splitter configuration with ``global'' spin orbit and single-moded
incoming leads.

\bibitem{BLS} G. Burkard \textit{et al.}, Phys.\ Rev.\ B \textbf{61}, R16303
(2000).

\bibitem{taddei} F. Taddei and R. Fazio, Phys.\ Rev.\ B \textbf{65}, 075317
(2002).

\bibitem{s-o} A. V. Moroz and C. H.\ W. Barnes, Phys.\ Rev. B \textbf{60},
14272 (1999), F. Mireles and G. Kirczenow, Phys. Rev. B \textbf{64}, 024426
(2001); and M. Governale and U. Z\"{u}licke, cond-mat/0201164.

\bibitem{apl-egues} J.\ C. Egues, G. Burkard, and D. Loss, to be submitted.

\bibitem{mireles} See, for instance, Mireles and Kirczenow in Ref. [%
\onlinecite{s-o}].

\bibitem{erasmo} E.\ A. de Andrada e Silva \emph{et al.} Phys. Rev. B
\textbf{55}, 16293 (1997). See also L. W. Molenkamp \textit{et al.} [Phys.
Rev. B \textbf{64}, R121202 (2001)] for a discussion of the relevance of the
proper matching of the velocity operator for transport across hybrid
ferromagnetic/semiconductor junctions.

\bibitem{am} N. W. Ashcroft and N. D. Mermin, \emph{Solid State Physics},
Ch. 9. (Holt, Rinehart, and Winston, New York, Chicago, etc., 1976).

\bibitem{buttiker} M. B\"{u}ttiker, Phys.\ Rev. B \textbf{46}, 12485 (1992);
Ya. M. Blanter and M. B\"{u}ttiker, Phys.\ Rep. \textbf{336}, 2 (2000).

\bibitem{molenkamp} L. W. Molenkamp \emph{et al.} in\ Ref. [%
\onlinecite{erasmo}] and M. H. Larsen \textit{et al}., cond-mat/0112175.

\bibitem{mismatch} Equation (\ref{eq12}) assumes unity transmission through
the Rashba region. The band-structure mismatch in our system is
solely due to the Rashba energy $\epsilon _{R}\equiv \hbar
^{2}k_{R}^{2}/2m^{\ast }$ [see Fig. 1(b)] and is extremelly small
since $\epsilon _{R}\ll \varepsilon _{F}$. We can estimate the
transmission amplitude from $t=2\left( 1+\epsilon _{R}/\varepsilon
_{F}\right) ^{1/4}/[1+(1+\epsilon _{R}/\varepsilon _{F})^{1/2}]$
(see Ref. [\onlinecite{molenkamp}]). For typical
parameters \cite{nitta} $\epsilon _{R}/\varepsilon _{F}=0.1$ we find $%
|t|^{2}=\allowbreak 0.99943.$

\bibitem{delay} The ``delay time'' $\tau _{d}$ between the injected partners
in an electron pair is assumed negligible as compared to the ``transit
time'' $\tau _{t}$ for an electron to cross the beam-splitter structure. For
typical parameters we find $\tau _{d}\sim 0.6$ ps and $\tau _{t}\sim 10-100$
ps (see Ref. [\onlinecite{egd}] for details).
\end{thebibliography}
\end{document}